# Metadata Enrichment of Long Text Documents using Large Language Models

**Lamba, Manika**  University of Oklahoma, USA | manika@ou.edu
**Peng, You**  University of Illinois Urbana-Champaign, USA | youpeng2@illinois.edu
**Nikolov, Sophie**  University of Illinois Urbana-Champaign, USA | nikolov3@illinois.edu
**Layne-Worthey, Glen**  University of Illinois Urbana-Champaign, USA | gworthey@illinois.edu
**Downie, J. Stephen**  University of Illinois Urbana-Champaign, USA | jdownie@illinois.edu

## ABSTRACT
In this project, we semantically enriched and enhanced the metadata of long text documents, theses and dissertations, retrieved from the HathiTrust Digital Library in English published from 1920 to 2020 through a combination of manual efforts and large language models. This dataset provides a valuable resource for advancing research in areas such as computational social science, digital humanities, and information science. Our paper shows that enriching metadata using LLMs is particularly beneficial for digital repositories by introducing additional metadata access points that may not have originally been foreseen to accommodate various content types. This approach is particularly effective for repositories that have significant missing data in their existing metadata fields, enhancing search results, and improving the accessibility of the digital repository.

## INTRODUCTION
Metadata quality is crucial for data discovery, accessibility, reproducibility, and interoperability (Liu et al., 2024; Alemneh, 2009; Park, 2009). Common metadata quality criteria include accuracy, completeness, and consistency (Park, 2009). In digital libraries, the combination of high metadata quality, effective metadata enhancement, and strategic metadata & semantic enrichment is critical for document organization and accessibility. Collectively, these features and processes not only enhance document discoverability but also support research activities, foster user engagement, and contribute to long-term data preservation. In recent years, metadata enrichment has been done by performing semantic-based tagging (Al-Natsheh et al., 2018), automatic extraction and semantic analysis (Bertin and Atanassova, 2012), and automatic keyword inference (Guo et al., 2013). Machine learning techniques have shown promise in extracting metadata from scientific articles (Tkaczyk, 2017; Tkaczyk et al., 2015). Moreover, recent research on metadata enhancement using large language models (LLMs) includes automated methods to detect and correct errors in electronic thesis & dissertation (ETD) metadata fields (Choudhury, et al., 2023). Despite progress, challenges remain in maintaining high-quality metadata due to the increasing volume of digital resources and the limitations of manual metadata creation (Alemneh, 2009; Hillmann, 2008). Previous papers have examined either manual or automated techniques for enhancing and enriching metadata. However, there is a significant gap in examining hybrid approaches: those that use manual enhancement and enrichment strategies with automated semantic strategies for improving metadata, specifically for long text documents such as theses and dissertations. Such an integrated approach could leverage the precision of human curation alongside the scalability of automated systems, potentially addressing both quality and volume challenges in metadata management of long text documents. In this case study, we introduce a hybrid approach of manual and automated methods for metadata enrichment and enhancement for thesis and dissertation metadata.

## METHODOLOGY
### Data Collection

For our study, we used the HathiTrust Digital Library (2025) to retrieve thesis and dissertation metadata records. We used the advanced search function in the HathiTrust catalog to do a comprehensive search for all ETDs that met our temporal and language requirements. During the search, we used keywords such as "dissertation", or "academic" in the subject field. In addition, we used the "Full View Only" option to ensure that the items we could add to the collection included full-text downloadable data. Finally, we created a collection of dissertation records and aggregated their HathiTrust identifiers (HTIDs) into a list. A total of 6,445 ETDs were identified spanning from 1920 to 2011 in English. From this collection, we formally applied to the HathiTrust Research Center team (https://analytics.hathitrust.org/) to access the full text of the ETDs. Once our request was approved, the data files were downloaded to our server for additional processing and analysis.

### Data Processing
The metadata for 6,445 ETDs was exported in JSON format from the HathiTrust digital repository. The JSON format has a complex structure, and although it is suitable for storing unstructured data, it is not easy to manipulate directly or use in the actual analysis process, especially when it comes to large-scale data cleaning and statistical analysis. In order to facilitate subsequent metadata enhancement and enrichment tasks, we converted the metadata file from JSON to CSV format. Additionally, the full-text data was fragmented, with each page of a specific ETD stored as an individual TXT file. This fragmentation added complexity to our data analysis process. We merged the multiple TXT files corresponding to the same ETD to address this issue into a single cohesive file.

## FINDINGS
### Metadata Enhancement and Enrichment

Metadata enhancement aims to refine existing metadata by correcting errors, standardizing formats, and aligning data with established metadata schemas or standards. This approach improves metadata consistency and interoperability across different systems, ensuring that users can access reliable and cohesive information across platforms. In contrast, metadata enrichment extends beyond basic refinement and adds new layers of information to metadata records to enhance their context and relevance. This process may include adding descriptors such as subject categories, keywords, summaries, or external linked data, and these enriched metadata records provide deeper context and increased usefulness for advanced searching and discovery. There were 28 pre-existing metadata fields associated with HathiTrust's ETD data: *htid, access, rights, ht_bib_key, description, source, source_bib_num, oclc_num, isbn, ISSN, lccn, title, imprint, rights_reason_code, rights_timestamp, us_gov_doc_flag, right_date_used, pub_place, lang, bib_fmt, collection_code, content_provider_code, responsible_entity_code, digitization_agent_code, access_profile_code, author, catalog_url, handle_url*. We found that many of these fields were not appropriate to describe the ETDs. Therefore, we selected 8 essential metadata fields, namely *htid, oclc-num, title, rights-date-used, collection-code, author, catalog-url, handle-url,* that were either specific to ETDs or required for document retrieval, and removed the rest.

Additionally, we used the Dublin Core metadata standard to rename the chosen field names (Table 1), as designed by the National Digital Library of Theses and Dissertations (NDLTD, 2023) for ETD data specifically. The *'author'* field in the original HathiTrust metadata consisted of the researcher's birth and death year in addition to their names, which we separated into two fields: *'dc:creator'* and *'dc:creator:dob'*. We added 10 more fields that were related to *advisor, committee chair, committee member, department, university, degree name, degree level, location, keywords, abstract*, all of which were absent in the HathiTrust metadata but which we deemed crucial for adequately describing ETDs data. The data for these additional fields, except for *keywords* and *abstract* fields, were manually added by accessing individual records through HathiTrust's catalog and handleURLs.

Table 1 shows the 20 metadata fields that we used to enrich and map the HathiTrust metadata to the Dublin Core standard for ETDs (NDLTD, 2023). In addition to enriching the ETD metadata, we conducted multiple rounds of manual data cleaning and normalization, ensuring consistency, accuracy, and usability. This meticulous process involved resolving records with multiple authors and degrees, standardizing name and date formats, and removing irrelevant and duplicate entries. Below are detailed examples of several challenges addressed during the cleaning process:

- **ETDs with Multiple Author Names:** ETDs with multiple authors were split into distinct rows for clarity. For example, for the ETD, *"The Nature of Gas-Metal Electrodes"*, co-authored by Sidney J. French and Louis Kahlenberg, a duplicate row was created, differing only by author name.
- **ETDs with Multiple Degrees:** ETDs with multiple degrees were separated into unique rows for each degree. For instance, Patrick Francis Quinn's ETD on *"The Fatalism of Herman Melville"* was listed under separate rows for "Bachelor of Arts" and "Master of Arts."
- **Multiple Authors earning Different Degrees for same ETD:** A more complex case involved ETDs with multiple authors earning different degree types. For ETDs co-authored by individuals earning different degrees, unique rows were created for each author-degree pair. For example, *"The Effect of Decreased Oxygen in the Respired Air Upon Metabolism, Acidosis, and the Differential Leukocyte Count"* involved five authors with varying degrees (e.g., Master of Science, Master of Arts, Bachelor of Science in Medical Sciences).
- **Author's Name Formatting:** Periods were removed from first names (e.g., "Margaret Anne." became "Margaret Anne").
- **Date of Birth Standardization**: Birth years formatted with a following hyphen (e.g., "1887-") were corrected to avoid confusion, resulting in entries like "1887."
- **Removing Duplicate and non-ETD Titles:** We excluded 1,131 entries misclassified as ETDs, along with duplicate titles.
- **De-duplication Complexity**: De-duplication was more complex than anticipated. There were several ETDs that were integrated back into the main metadata file because of version errors and discrepancies between the *'dc:identifier: catalog'* field versus the *'dc:identifier:handle'* field. In these cases, *'handle-url'* showed a version of the ETD without a title page whereas the *'catalog-url'* field for the same ETD, provided multiple versions of the ETD where at least one of those versions provided a title page, which was helpful for adding more metadata.

| Old Field Name | New Field Name | Description |
|---|---|---|
| htid | dc:identifier:htid | Unique ID number assigned to HathiTrust volumes |
| oclc-num | dc:identifier:oclc | OCLC control number for the item |
| collection-code | dc:relation:isPartOf | Code identifying the specific collection within HathiTrust |
| title | dc:title | The title of the ETD |
| author | dc:creator | Name of the author who wrote the ETD |
| author | dc:creator:dob | Author's date of birth and date of death, if applicable |
| rights_date_used | dc:date:issued | Publication year |
| n/a | dc:contributor:advisor | Faculty member who served as the primary advisor |
| n/a | dc:contributor:committeeChair | Faculty member who served as the chair |
| n/a | dc:contributor:committeeMember | Faculty members who served on the committee |
| n/a | thesis:degree:department | Academic department associated with the program |
| n/a | thesis:degree:discipline | Academic field or discipline of study |
| n/a | thesis:degree:grantor | Institution awarding the degree |
| n/a | thesis:degree:name | Name of the degree awarded (e.g., Ph.D., M.S., Ed.D.) |
| n/a | thesis:degree:level | Level of the degree (e.g., Doctoral, Master's) |
| n/a | dc:coverage:spatial | Location of the university |
| n/a | dc:subject | Keywords or subject terms describing the thesis |
| n/a | dc:description:abstract | Summary or abstract of the thesis |
| catalog-url | dc:identifier:catalog | Permanent URL or identifier for the catalog record |
| handle-url | dc:identifier:handle | The direct URL link to the ETD provided by HathiTrust |

*Note: n/a represents that those fields were absent in the HathiTrust metadata*

**Table 1. Metadata Enrichment for HathiTrust ETDs using Dublin Core Fields**

Despite our best efforts to manually complete the metadata by reviewing each ETD's title and preliminary pages, many fields remain empty and are marked as '*NA*' in the dataset. These fields include *'dc:contributor:advisor'*, *'dc:contributor:committeeChair'*, *'dc:contributor:committeeMember'*, *'thesis:degree:department',* and *'thesis:degree:discipline'.* This is because we could not find the relevant information from the title or preliminary pages of the ETDs available in the HathiTrust repository to populate these fields. After removing duplicates and non-ETD titles, our final dataset contained metadata for 5,760 ETDs. There were issues identified during the data processing of ETDs and which were removed from the final dataset. A total of 1,131 titles were mis-classified as ETDs. Additionally, 216 ETDs contained poor-quality text due to Optical Character Recognition (OCR) errors, and another 56 titles lacked availability of full-text files.

**Semantic Enrichment**

Semantic enrichment is slightly different than metadata enrichment as it adds meaning and contextual connections to the document's content. It involves understanding the content at a deeper level (e.g., themes, entities, relationships) and linking it with relevant data sources, entities, or concepts. Semantic enrichment goes beyond mere tagging to enhance comprehension and context, often relying on ontologies or knowledge graphs to relate data points (Smith, 2024). To enrich the metadata semantically, we applied keyword extraction and text summarization techniques using LLMs. To ensure the quality and accuracy of the LLM-generated results, we randomly selected and manually reviewed 10% of the metadata records enriched through keyword extraction and text summarization techniques against the original source documents, confirming their accuracy.

*Keyword Extraction*

KeyLLM (2024) is a keyword extraction algorithm that uses LLM to infer keywords from the text (Fig. 1a). It can run independently or in combination with KeyBERT to improve both efficiency and accuracy. By utilizing the capabilities of LLMs, KeyLLM detects not just 'keywords' as a list of tokens, but also the key themes of the document if they are not mentioned. Alternatively, it can be used to perform simple keyword detection. This algorithm is flexible and fully compatible with OpenAI's LLMs to provide customized keyword-extraction solutions. In our study, we used the *'title'* field to extract relevant keywords using KeyLLM and populated the results in the *'dc:subject'* metadata field.

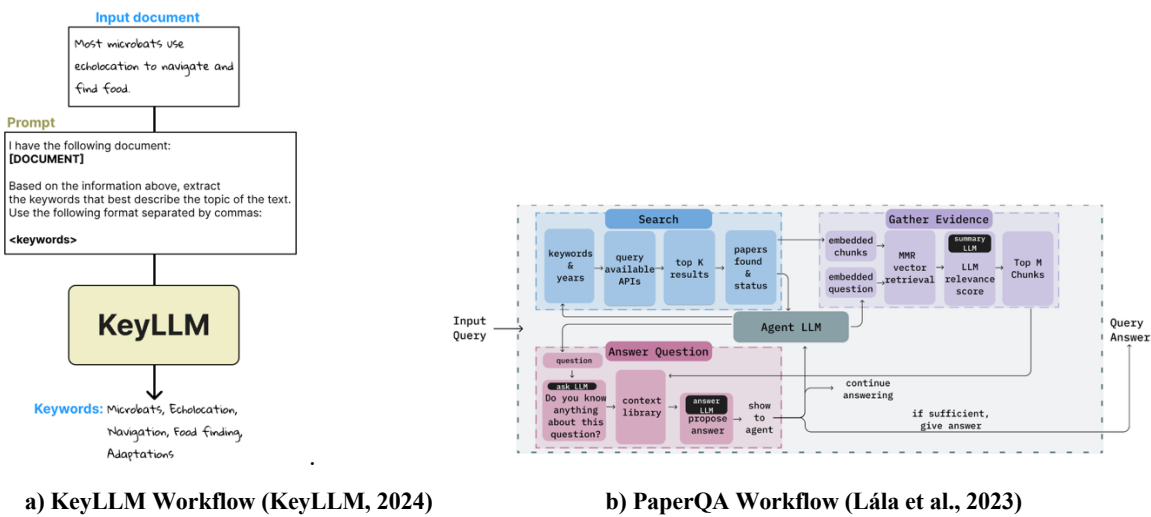

a) KeyLLM Workflow (KeyLLM, 2024)          b) PaperQA Workflow (Lála et al., 2023)

**Figure 1. Workflows used for Semantic Enrichment**

*Text Summarization*

Paper-QA (Lála et al., 2023) is a high-precision Q&A and abstract generation package specialized for scientific documents (Fig. 1b). It utilizes the Retrieval Augmented Generation (RAG) model to help users efficiently obtain answers from scientific documents by indexing PDFs or text files and generating answers with in-text citations. Paper-QA searches, retrieves, and summarizes scientific documents with the aid of LLMs, while incorporating metadata to increase the relevance and accuracy of the answers. For our study, we used the Paper-QA package (https://pypi.org/project/paper-qa/) combined with the OpenAI GPT-4o-mini model (https://openai.com/index/gpt-4o-mini-advancing-cost-efficient-intelligence/) to generate coherent and comprehensive abstracts from full-text version of 5,760 ETDs which were then used to populate the *'dc:description:abstract'* field of the metadata. To ensure the completeness and accuracy of the input data, we first merged each document's segmented text files (originally separated into pages) into a single file for each unique *'htid'* identifier. Each document file was placed into an *'htid-named'* directory, thereby preserving the content's continuity and optimizing it for input into the summarization model.

Following the data preparation, we exported metadata from the previous metadata enrichment section as a CSV file, using the *'htid'* as an index to systematically iterate through each record. For each document, we employed the Paper-QA tool to prompt the model with the question: *"What is the abstract of this paper?"* This method facilitated the generation of structured summaries, each output containing clearly defined *"Question," "Answer,"* and *"Reference"* components. The summaries were subsequently saved as .txt files for further usage. Then we used a Python script to extract those abstracts generated by the Paper-QA tool. By leveraging the *os* and *pandas* libraries, the script efficiently located each document's corresponding .txt file, checked for content availability, and extracted only the essential *"Answer"* section while excluding auxiliary text. This script could get three different results. The first outcome was a successfully generated abstract, indicating that both the metadata and full text were provided, allowing the model to summarize the document accurately. The second outcome, *"File does not exist,"* meant that although the HathiTrust repository provided the document's metadata, the full-text file itself was not available. The third outcome, *"I cannot answer"*, reflected instances where poor OCR quality or other errors in the full-text prevented the generation of a coherent abstract. This structured approach ensured the efficient handling of data variations, providing clear insights into document availability and quality issues across the dataset.

**CONCLUSION**

In this paper, we present a dataset of theses & dissertations retrieved from the HathiTrust digital repository. This metadata was semantically enriched and enhanced using manual and automated efforts using LLMs. This metadata standardization serves two key purposes: (i) it ensures that ETD metadata follows established community standards; and (ii) it enhances interoperability, enabling seamless metadata exchange between HathiTrust and other institutional repositories that adhere to Dublin Core standards. Furthermore, the automated metadata enrichment process using LLMs helped us include two additional fields, keywords and abstracts, that were not part of the original HathiTrust dataset. This approach introduces additional metadata access points for digital repositories, which may not have been originally designed to accommodate other types of content. For example, HathiTrust's metadata is designed primarily for books rather than theses and dissertations. Therefore, the access points provided by HathiTrust are not wholly suitable for ETDs. The methodology of semantic enrichment using LLMs discussed in this paper addresses this limitation by making it possible to generate and integrate missing metadata fields with a high degree of semantic accuracy. The methodology presented in this paper extend beyond this single case study and has can be replicated across other digital libraries and databases that face similar challenges with incomplete or inadequate ETD metadata. Many institutional repositories worldwide struggle with inconsistent or missing metadata fields for their thesis and dissertation collections, and the hybrid approach demonstrated in this work could provide a scalable solution for addressing these gaps.

## DATA AVAILABILITY

The dataset introduced and analyzed in this paper is available for download and reuse under a CC BY-NC-SA 4.0 license in Zenodo at https://doi.org/10.5281/zenodo.14172717. A citation to the dataset is provided in the reference list, as Lamba et al. (2024).

## GENERATIVE AI USE

We employed OpenAI's GPT-4o-mini model for semantic enrichment as described in the paper. We evaluated the model's output by human review. The authors assume all responsibility for the content of this submission.

## AUTHOR ATTRIBUTION

**Lamba**: conceptualization, methodology, investigation, writing – original draft, writing – review and Editing, supervision; **Peng and Nikolov**: data curation, formal analysis, visualization; **Layne-Worthey and Downie**: project administration, writing – review and editing.

## ACKNOWLEDGMENTS

We sincerely thank the HathiTrust Research Center Services team for their assistance in providing the data.